\documentclass[twocolumn,aps,superscriptaddress]{revtex4}
\usepackage{latexsym,amssymb,amsmath}
\usepackage{color}
\usepackage[bookmarksnumbered,bookmarksopen,colorlinks,citecolor=red,linkcolor=blue]{hyperref}
\usepackage{mathrsfs}
\usepackage{times}
\usepackage{csquotes}
\usepackage{amssymb}
\usepackage{amsmath}
\usepackage{graphicx}
\usepackage{subfig}
\usepackage[normalem]{ulem}
\usepackage{makecell}
\usepackage{multirow}
\usepackage{color}

\begin{document}

\title{Machine learning transforms the inference of the nuclear equation of state}

\author{Yongjia Wang}
\affiliation{School of Science, Huzhou University, Huzhou 313000, China}

\author{Qingfeng Li\footnote{Corresponding author: liqf@zjhu.edu.cn}}
\affiliation{School of Science, Huzhou University, Huzhou 313000,  China}
\affiliation{Institute of Modern Physics, Chinese Academy of Sciences, Lanzhou 730000, China}

%




\maketitle
The field of nuclear physics relies on experiments and theoretical models to understand the structure and properties of nuclei and their reactions in order to bring insights into the nuclear
forces that holds protons and neutrons together in an atomic nucleus. The nuclear equation of state (EOS) usually describes the thermodynamic relationship between the binding energy per nucleon $E$ and density $\rho$, as
well as the isospin asymmetry $\delta$ in
nuclear matter (ideally, a uniform and infinite system with neutrons and protons), it also reveals rich information about the nature of nuclear forces in the nuclear medium. The knowledge of the nuclear EOS is essential for our understanding of the properties and structures of both nuclei and neutron stars, the dynamics of heavy ion collisions
(HICs), as well as supernova explosions and neutron-star mergers. Thus, the nuclear EOS has attracted considerable attention
from both nuclear physics and astrophysics communities
since long time ago \cite{drischler2021chiral,li2008recent,oertel2017equations,sorensen2023dense,lattimer2021neutron,Lopez:2020zne}. From experimental point of view, the nuclear EOS cannot be directly measured in terrestrial laboratory because the created nuclear matter in HICs exists only for a very short time (typically several fm$/$c). From theoretical point of view, the calculation of EOS relies heavily on microscopic nuclear many-body theory and nuclear force. Neither of them are completely understood, although decades of efforts have been devoted to study them, and great progress has been made \cite{tsang2012constraints,oertel2017equations}. Alternatively, the nuclear EOS can be extracted or inferred from measurements on nuclear structure, astrophysical observations of neutron stars, and observables from HICs \cite{WOLTER2022103962,Wang:2020vwb}.

To quantitatively describe the nuclear EOS, the following expression is frequently used:
$E(\rho,\delta)=E(\rho,\delta=0) + E_{\rm sym}(\rho)\delta^{2} + \mathcal{O}(\delta^{4})$.
The first term $E(\rho,\delta=0)$ is the binding energy per nucleon in the isospin-symmetric nuclear matter, $E_{\rm sym}(\rho)$ is the density-dependent nuclear symmetry energy. Investigating how $E(\rho,\delta=0)$ and $E_{\rm sym}(\rho)$ vary with density, is one of the scientific goals of the current and upcoming nuclear facilities (e.g., the CSR and HIAF in China, the FRIB in the United States, the RIBF in Japan, the SPIRAL2 in France, the FAIR in Germany, and the NICA in Russia) around the world. Great endeavors have been made, however a precise picture of the nuclear EOS in a broad
range of densities and isospin asymmetries has still not been obtained \cite{li2008recent,tsang2012constraints,oertel2017equations}.

The inference of the nuclear EOS usually relies on data from both experimental measurements and theoretical calculations, which is similar to the process of mining information from the Big Data. Machine learning (ML) algorithm
has a strong capability to infer and mine information from data. Indeed, ML has been a branch of artificial intelligence and computer science since 1980s, and mainly focuses on the use of data and algorithms to imitate the way that humans learn based on various mathematical models \cite{jordan2015machine}. Thanks to the appearance of many modern and highly
integrated programming tools, high performance personal computers and open-source tools and platforms, ML or deep learning (which is a subset of ML but with a deeper and more complex mathematical structure) has become very
popular in many fields of natural science (such as physics, see Ref\cite{carleo2019machine} for recent review),
and brings the fourth paradigm of science, which is so-called data-driven discovery \cite{hey2009fourth}.

\begin{figure*}
    \centering
    \includegraphics[width=0.9\textwidth]{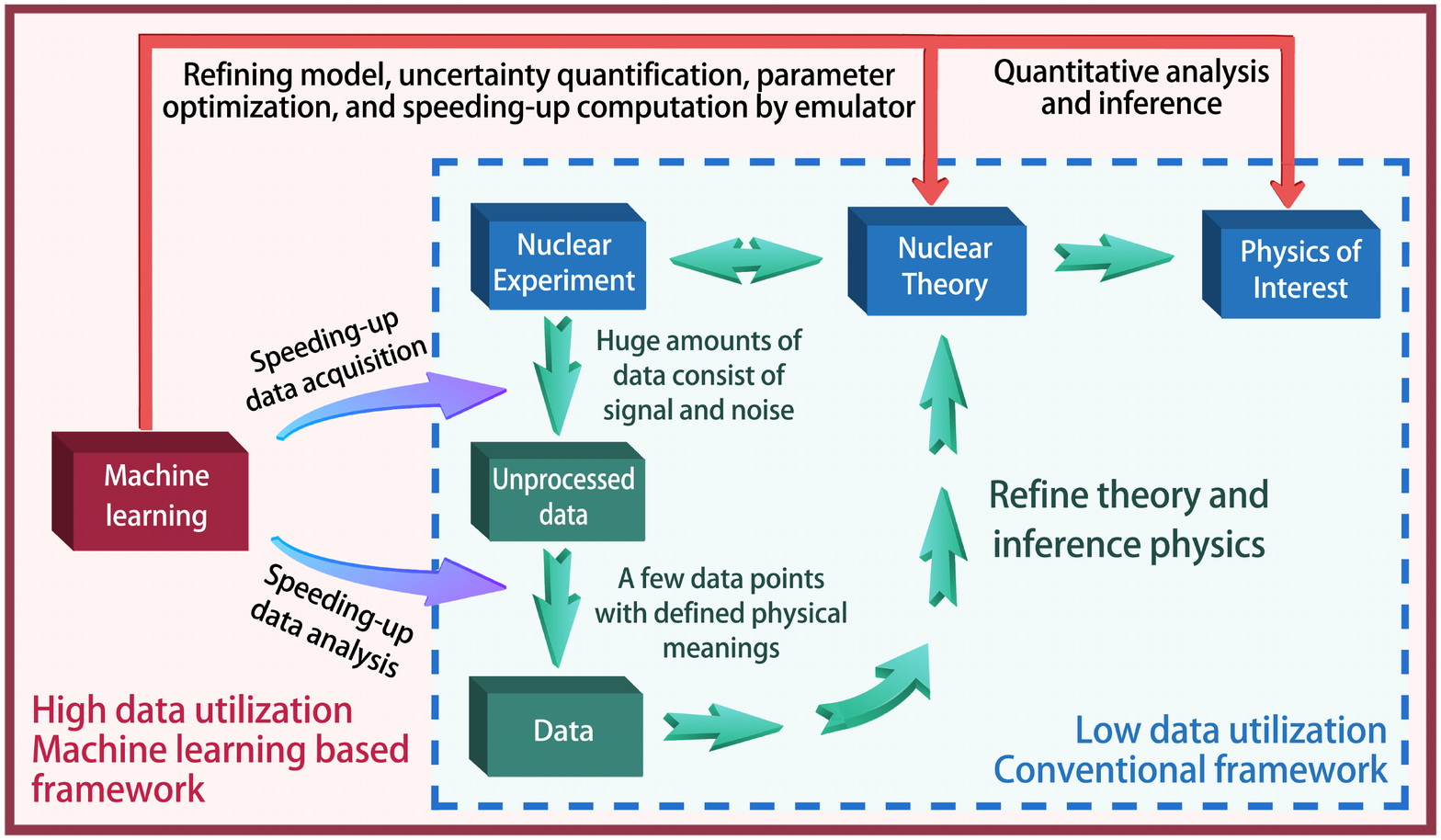}
    \caption{Conventional framework (dashed-line box) and machine learning based framework (solid-line box) in the study of nuclear physics.}
    \label{fig:1}
\end{figure*}
 In the conventional framework for studying nuclear physics, usually huge amounts of data that consist of both signal (which is the information one wants to obtain) and noise (which is something one wishes to forgo) are acquired from nuclear experiments. Then, the acquired big data will be processed and analyzed in order to obtain observables with defined physical meanings according to the prior human knowledge. These observables usually only have a few data points and will be used to infer physics of interest with the combination of nuclear theory. A well-known example of the conventional framework in nuclear physics is the studying of nuclear EOS with HICs. The size of data acquired per second from detectors in HIC experiments can reach as high as several terabytes, and these data will be processed to obtain observables, such as the collective flows which are only a few data points. The information of nuclear EOS is then extracted from the comparison of the collective flows obtained with transport model simulations to those with experimental
measurements. Apparently, the extent of data used in the conventional framework is very limited.

The application of ML has changed the way we study nuclear physics.
Indeed, pioneering work on studying nuclear physics with ML started in the early 1990s, where artificial neural networks (ANN) were used to study global nuclear properties \cite{gazula1992learning}. Later on, Bass et al. demonstrated that ANN is capable of improving the accuracy of the impact parameter determination in HICs \cite{bass1994neural}, which is the pioneering application of ML in heavy-ion physics. Since then such studies are continually performed and are rapidly becoming popular research topics especially in the recent decade \cite{bedaque2021ai,boehnlein2022colloquium,he2023high,he2023machine}. As illustrated in Fig.\ref{fig:1}, ML can alter the full chain of the conventional framework. For example, it has been found that ML not only can speeding-up data acquisition and analysis in nuclear experiments, but also can refine model, accelerate computation by emulators, optimize model parameters and quantify uncertainties in nuclear theory. For example, various ML algorithms have been shown to be rather powerful in improving the performance of nuclear models on the prediction of nuclear masses, nuclear charged radii, half-lives, and so on. See, e.g., Refs \cite{bedaque2021ai,boehnlein2022colloquium,he2023high,he2023machine} for recent reviews.
In addition, quantitative analysis methods, such as Bayesian inference, are also widely used to infer physical quantities of interest from the comparison of observables determined experimentally with that obtained from theoretical model calculations \cite{morfouace2019constraining,man2022,baoan2023}. Extraction of the nuclear EOS is ideally suited for the Bayesian inference. According to the Bayes's theorem, constraints on a physical quantity $\theta$ (or vector of parameters) from a set of experimental measurements $\mathbf{y_{exp}}$ are described by the posterior probability $P(\theta|\mathbf{y_{exp}})$ and can be written as:
\begin{equation}
P(\theta|\mathbf{y_{exp}}) \propto P(\mathbf{y_{exp}}|\theta) P(\theta).\qquad\qquad\qquad\qquad\qquad
\end{equation}
Here the prior $P(\theta)$ reflects any preexisting information/belief about the distribution of the physical quantity $\theta$. $P(\mathbf{y_{exp}}|\theta)$ is likelihood function that denotes the probability of a theoretical model being
matched with the data $\mathbf{y_{exp}}$ at a given value of its parameter $\theta$. Both  experimental and theoretical model uncertainties can be considered in the construction of $P(\mathbf{y_{exp}}|\theta)$. Usually, it requires massive computational work of running theoretical model with different choices of $\theta$, especially when $\theta$ is a vector of parameters (which is common in nuclear physics), so, this is time consuming and not feasible until the last decade. In a work recently reported in Nature, Huth et al. extracted the EOS of dense nuclear matter by using Bayesian inference which combines data from HICs and astrophysical
observations with microscopic nuclear theory calculations \cite{huth2022constraining}. This work is novel in that it demonstrates how the interdisciplinary analyses can shed light on the nuclear EOS, and shows that EOS constraints from
HIC experiments and multi-messenger astrophysics are remarkably consistent with each other, which provides important information on the nuclear EOS at supra-saturation densities.

Besides Bayesian inference, another emblematic example, for which modern ML methodology is also ideally suited, is the use of various supervised machine learning algorithms to extract information from event-by-event HIC data. The application of ML in heavy-ion physics is expected to be promising because there is a huge amount of event-by-event data either from experimental measurements or from theoretical model simulations. Event-by-event data in HICs encodes rich information about relevant physics and correlations
among observables which can be sensitive to the physics of interest but independent of other model parameters. With the help of ML algorithm, physical information can be extracted from observables in HICs
on the event-by-event analysis, see e.g., Refs \cite{pang2018equation,wang2022decoding}, instead of traditional method which usually relies on the expectation values
of observables over all considered events. As demonstrated in our recent work \cite{wang2022decoding}, by using a modern ML algorithm and data generated with a microscopic transport model, we presented a ML based framework for inferring the
$E_{\rm sym}(\rho)$ from observables in HICs on the event-by-event analysis, and the
framework is found to be robust against variations in model parameters. Note that unlike the conventional framework in which observables are usually constructed according to the prior human knowledge and their number is limited, observables (or features) in the ML based framework can be constructed by ML algorithms themselves without any prior knowledge from human and their number can be infinite in general. Apparently, the extent of data used in the ML based framework is much higher than that in conventional framework.
Moreover, the most important observables that drive ML predictions can be identified by various feature attribution methods, thereby offering valuable insights on the physics of interest. This ML based framework as a data-driven methodology, is extensible and may offer a new paradigm to study other issues in HICs.
Besides studying the nuclear EOS, various ML algorithms have been used to study diverse topics in HICs from low and intermediate energies up to relativistic energies, such as, the impact parameter, the nuclear liquid-gas phase transition, initial state nuclear structure, and QCD phase transition. See, e.g., Refs \cite{boehnlein2022colloquium,he2023high,he2023machine,zhou2023} for recent reviews and comprehensive lists of references.

The marriage of ML and heavy-ion physics brings a ML based framework which can be a powerful tool and may open a new venue to study the underlying physics in HICs, and has revolutionized the way we infer the nuclear EOS. In the recent years, the number
of reported applications of ML is growing at an extraordinary rate, several promising results have been shown, but we are still in the earliest stages of exploring the application of ML in the study of heavy-ion physics. Many open questions remain, such as, whether the output of ML is interpretable or explainable (the interpretability or explainability, it is the degree to which a human can understand the cause of a decision. This is important, particularly for physicists, because understanding what happens when
ML algorithms make predictions could help us make better
use of the outputs), whether ML algorithms have the ability to infer information from vastly different data sets (generalizability, it refers to the capacity of the trained ML algorithms to adequately predict on data which sources differ from those involved in training. ML algorithms which have high generalizability are usually preferred, especially when the data are generated with models.), whether knowledge can be learned from small datasets (data-hungry, it matters especially when the cost of data generation is high).

Within the next decade, new heavy-ion facilities, such as HIAF, NICA, and FAIR, will generate data at far greater rates than the present facilities. The interpretation of these big data will become even more challenging.
Fortunately, ML is one of today's most rapidly growing technical fields and continuously producing tools
that are potentially applicable to a wide array of tasks in heavy-ion physics. Moving forward, the mergence of ML and heavy-ion physics can be improved in at least four aspects. First, improving the quality of the generated synthetic training data. Because real data do not come with labels, model simulations provide the only possible way to get data sets have truth labels for training. More sophisticated models are expected to be more appropriate to generate synthetic data that can closely mimic the real data.  Alternatively, using data that generated from different models may be of benefit to achieve an model-independent extraction of the nuclear EOS with ML algorithms. Second, introducing physical information into ML algorithms. This can be done either by using input features with defined physical meanings or by considering physical symmetries and laws when constructing architectures of ML algorithms \cite{raissi2019physics}. Third, using successful experiences of ML applications in other fields, such as condensed matter physics and particle physics. The use of ML in those fields  is more advanced than in heavy-ion physics, many promising techniques can be extended to study heavy-ion physics. Fourth, changing the latest developments in ML into tools for studying heavy-ion physics. ML is a rapidly growing and flourishing field fueled by the successful applications in many aspects of our daily life. Nowadays, a diverse array of ML algorithm has been developed and continue to be refined to cover a wide variety of data types and tasks, this is a sufficiently large and diverse pool of tools feasible to study heavy-ion physics.

%
The work is supported in part by the National Natural Science Foundation of China (Nos. U2032145 and 11875125), the National Key Research and Development Program of China under Grant No. 2020YFE0202002.

\bibliography{apssamp}

\begin{thebibliography}{10}

\bibitem{drischler2021chiral}
C.~Drischler, J.~Holt, C.~Wellenhofer, {\it Annual Review of Nuclear and
  Particle Science\/} {\bf 71}, 403 (2021).

\bibitem{li2008recent}
B.-A. Li, L.-W. Chen, C.~M. Ko, {\it Physics Reports\/} {\bf 464}, 113 (2008).

\bibitem{oertel2017equations}
M.~Oertel, M.~Hempel, T.~Kl{\"a}hn, {\it et~al.\/}, {\it Reviews of Modern
  Physics\/} {\bf 89}, 015007 (2017).

\bibitem{sorensen2023dense}
A.~Sorensen, K.~Agarwal, K.~W. Brown, {\it et~al.\/}, {\it arXiv preprint
  arXiv:2301.13253\/}  (2023).

\bibitem{lattimer2021neutron}
J.~Lattimer, {\it Annual Review of Nuclear and Particle Science\/} {\bf 71},
  433 (2021).

\bibitem{Lopez:2020zne}
J.~A. Lopez, C.~O. Dorso, G.~A. Frank, {\it Front. Phys. (Beijing)\/} {\bf 16},
  24301 (2021).

\bibitem{tsang2012constraints}
M.~Tsang, J.~Stone, F.~Camera, {\it et~al.\/}, {\it Physical Review C\/} {\bf
  86}, 015803 (2012).

\bibitem{WOLTER2022103962}
H.~{Wolter}, M.~{Colonna}, D.~{Cozma}, {\it et~al.\/}, {\it Progress in
  Particle and Nuclear Physics\/} {\bf 125}, 103962 (2022).

\bibitem{Wang:2020vwb}
Y.-J. Wang, Q.-F. Li, {\it Front. Phys. (Beijing)\/} {\bf 15}, 44302 (2020).

\bibitem{jordan2015machine}
M.~I. Jordan, T.~M. Mitchell, {\it Science\/} {\bf 349}, 255 (2015).

\bibitem{carleo2019machine}
G.~Carleo, I.~Cirac, K.~Cranmer, {\it et~al.\/}, {\it Reviews of Modern
  Physics\/} {\bf 91}, 045002 (2019).

\bibitem{hey2009fourth}
A.~J. Hey, S.~Tansley, K.~M. Tolle, {\it et~al.\/}, {\it The fourth paradigm:
  data-intensive scientific discovery\/}, vol.~1 (Microsoft research Redmond,
  WA, 2009).

\bibitem{gazula1992learning}
S.~Gazula, J.~Clark, H.~Bohr, {\it Nuclear Physics A\/} {\bf 540}, 1 (1992).

\bibitem{bass1994neural}
S.~Bass, A.~Bischoff, C.~Hartnack, {\it et~al.\/}, {\it Journal of Physics G:
  Nuclear and Particle Physics\/} {\bf 20}, L21 (1994).

\bibitem{bedaque2021ai}
P.~Bedaque, A.~Boehnlein, M.~Cromaz, {\it et~al.\/}, {\it The European Physical
  Journal A\/} {\bf 57}, 1 (2021).

\bibitem{boehnlein2022colloquium}
A.~Boehnlein, M.~Diefenthaler, N.~Sato, {\it et~al.\/}, {\it Reviews of Modern
  Physics\/} {\bf 94}, 031003 (2022).

\bibitem{he2023high}
W.-B. He, Y.-G. Ma, L.-G. Pang, {\it et~al.\/}, {\it arXiv preprint
  arXiv:2303.06752\/}  (2023).

\bibitem{he2023machine}
W.~He, Q.~Li, Y.~Ma, {\it et~al.\/}, {\it arXiv preprint arXiv:2301.06396\/}
  (2023).

\bibitem{morfouace2019constraining}
P.~Morfouace, C.~Tsang, Y.~Zhang, {\it et~al.\/}, {\it Physics Letters B\/}
  {\bf 799}, 135045 (2019).

\bibitem{man2022}
M.~O. Kuttan, J.~Steinheimer, K.~Zhou, {\it et~al.\/}, {\it arXiv preprint
  arXiv:2211.11670\/}  (2022).

\bibitem{baoan2023}
B.-A. Li, W.-J. Xie, {\it arXiv preprint arXiv:2303.10474\/}  (2023).

\bibitem{huth2022constraining}
S.~Huth, P.~T. Pang, I.~Tews, {\it et~al.\/}, {\it Nature\/} {\bf 606}, 276
  (2022).

\bibitem{pang2018equation}
L.-G. Pang, K.~Zhou, N.~Su, {\it et~al.\/}, {\it Nature communications\/} {\bf
  9}, 210 (2018).

\bibitem{wang2022decoding}
Y.~Wang, Z.~Gao, H.~L{\"u}, {\it et~al.\/}, {\it Physics Letters B\/} {\bf
  835}, 137508 (2022).

\bibitem{zhou2023}
K.~Zhou, L.~Wang, L.-G. Pang, {\it et~al.\/}, {\it arXiv preprint
  arXiv:2303.15136\/}  (2023).

\bibitem{raissi2019physics}
M.~Raissi, P.~Perdikaris, G.~E. Karniadakis, {\it Journal of Computational
  physics\/} {\bf 378}, 686 (2019).

\end{thebibliography}
\bibliographystyle{Science}

\end{document}